\documentclass[aps,prb,manuscript]{revtex4}
\usepackage{epsfig}
\begin{document}
\title{Quadratic operators used in deducing exact ground states for correlated
systems: ferromagnetism at half filling provided by a dispersive band.}
\author{
Istv\'an~Chalupa and Zsolt~Gul\'acsi}
\affiliation{
Department of Theoretical Physics, University of Debrecen,
H-4010 Debrecen, Hungary.}
\date{\today}
\begin{abstract}
Quadratic operators are used in transforming the model Hamiltonian $\hat H$ of
one correlated and dispersive band in an unique positive semidefinite form
coopting both the kinetic and interacting part of the $\hat H$. The expression
is used in deducing exact ground states which are minimum energy eigenstates 
only of the full Hamiltonian. 
It is shown in this frame that at half filling, also dispersive bands can 
provide ferromagnetism in exact terms by correlation effects . 
\end{abstract}
\maketitle

\section{Introduction}

Exact results for non-integrable many body models open new perspectives in the 
description and understanding of physical systems of large interest, or
physical situations capturing attention by unusual properties. Metallic 
ferromagnetism \cite{int1}, superconductivity in ladder systems \cite{int2},
exotic phases in polymer chains \cite{int3}, subtle conducting and insulating 
phases in rare earth compounds \cite{int4}, 
spontaneous ferroelectricity \cite{int5}, stripes or checkerboards in two
dimensions \cite{int6}, or metal-insulator transition in two dimensional 
disordered systems \cite{int7} are some examples in this respect. 

On the technical side this field is in fact almost completely open, hence 
several procedures have been tested or are under development. Without to be
exhaustive, we exemplify below some of these, as for example 
process definition 
followed by unapproximated treatment of the model conditions \cite{el1},
Hilbert space properties captured in the low density limit \cite{el2}, 
equal rigorous upper and lower bounds for the ground state energy \cite{el3},
unitary transformation of eigenfunctions and Hamiltonians \cite{el4}, 
Schrieffer-Wolff transformations effectuated up to infinite order \cite{el5},
the optimal ground state method \cite{el6}, or the use of positive 
semidefinite operators and related techniques \cite{el7}. In the present paper 
we concentrate on, and develop the last of these methods since the procedure 
based on positive semidefinite operators often penetrates in other techniques 
as well.

By using positive semidefinite operators ($\hat P)$ in deducing exact ground 
states ($|\Psi_g\rangle)$, one first casts the Hamiltonian $(\hat H)$ in 
a positive semidefinite form $\hat H=\hat P +C$, where 
$C$ is a constant depending on the coupling constants of the starting 
$\hat H$. Since zero is the minimum possible eigenvalue of $\hat P$, 
the unapproximated 
ground state is obtained from the requirement $\hat P |\Psi_g\rangle=0$.
In the process of transforming the starting Hamiltonian in a positive 
semidefinite form, mostly one finds or uses 
$\hat P = \sum_{n}^{n_{max}} \hat P_n$, 
where the number $n_{max}$ of different types of positive semidefinite
terms often can reach even $n_{max}=8-10$, their structure being diverse,
containing linear, bi-linear, cubic or quartic combinations of the starting 
fermionic operators \cite{el8a}. The requirement for the ground state 
in the presence of different $\hat P_n$ contributions 
becomes $\hat P_n |\Psi_g\rangle=0$ for all $n \in [1,n_{max}]$, hence the 
deduced $|\Psi_g\rangle$ represents separately the ground state of different
parts of the Hamiltonian, often being the ground state separately of
the kinetic and interacting part of $\hat H$, which reduces 
considerably the application possibilities of the method. 

In this paper we overcome this inconvenience by using quadratic, e.g.
nonlinear combinations of the starting fermionic operators in defining 
new non-fermionic
operators $\hat A_{\bf i}$. Here the index ${\bf i}$ represents a lattice site,
but $\hat A_{\bf i}$ is not local, has a finite extension since captures 
contributions also from the neighborhood of the site ${\bf i}$.
The operators $\hat A_{\bf i}$ are used in the construction of an unique
$\hat G =\sum_{\bf i}\hat A^{\dagger}_{{\bf i}} \hat A_{\bf i}$ positive 
semidefinite form, leading to the expression of the transformed Hamiltonian
$\hat H= \hat G +C$, containing an unique positive semidefinite 
operator, namely
$\hat G$ (the contribution $C$, as before, being a constant). Excepting the 
test of this procedure taken at infinitely 
large on-site Coulomb repulsion \cite{el8}, for finite and nonzero value of 
the interaction this approach is for the first time applied here 
in this paper. The gain one achieves is that now an unique
positive semidefinite operator concentrates both the kinetic and
interaction part of $\hat H$. Hence the corresponding ground states are not
independently eigenstates of the kinetic and interacting part, or 
independently eigenstates of different parts and terms from $\hat H$,
but only eigenstates with minimal energy of the full Hamiltonian. 

Relating the studied Hamiltonian, one concentrates here on one-band correlated 
itinerant systems holding as well spin-spin interactions. The model 
Hamiltonian is in this case of extended Hubbard type (denoted hereafter by
$\hat H_0$) containing also spin-spin interactions $\hat H_S$, hence
$\hat H = \hat H_0 + \hat H_S$ holds. A such Hamiltonian structure is 
particularly often debated in connection to Kondo lattices \cite{el9} 
or $t-J$ type of models \cite{el10}, but generally emerges in the study of
strong correlation effects when additional interactions between nearest 
neighbours are taken into consideration\cite{el11}, or real materials 
\cite{el12,el12a} are described. We further note that in the presence of 
strong correlations, similar models have been also
used in the study of charge-ordered states, and metal-insulator transitions
in non-ferromagnetic cases \cite{vonsovski}. 

The here deduced exact eigenstates characterize dispersive bands at 
half filling 
and are localized and fully saturated ferromagnetic ground states only of
the full $\hat H$. At this point we remark that the exact ferromagnetic
ground states at finite concentration of carriers deduced up to this moment
are or related to a specific band structure (flat band 
ferromagnetism \cite{f1}, or Lieb's degenerate middle band ferromagnetism in a 
bipartite lattice \cite{f2}), or appear given by the ability of the
interaction to select the ground state from the existing minimum energy
eigenstates of the kinetic (not necessarily one-particle \cite{f2a}) part
\cite{el8a}. Our results show that arbitrary dispersive half filled bands can
be ferromagnetic in exact terms. Since only the whole Hamiltonian provides
these ground states, these not emerge given by a specific band structure,
nor appear as interaction selected ground states from existent eigenstates
of other parts of the Hamiltonian, but are created by both kinetic and 
interaction contributions, hence their eigenstate nature disappears when
the interactions are turned off, or when the kinetic part is neglected (e.g. 
in the localized limit).

The remaining part of the paper is structured as follows. Section II.
describes the model used, Section III. presents the exact transformation of 
the Hamiltonian in positive semidefinite form, Section IV. gives the 
general expression of the deduced ground states and presents the proof of the
ground state nature, Section V. describes particular cases when the deduced
ground states hold and analyzes the physical properties of the solutions.
Finally, Section VI. concluding the paper, closes the presentation.

\section{The used Hamiltonian}

Our starting Hamiltonian $\hat H=\hat H_0 + \hat H_S$ describes in one 
dimension a correlated band characterized by 
\begin{eqnarray}
\hat H_0 &=& \sum_{{\bf i},\sigma} (t_{\sigma} \hat c^{\dagger}_{{\bf i},
\sigma} \hat c_{{\bf i}+1,\sigma} + H.c.) + 
U \sum_{\bf i} \hat n_{{\bf i},\uparrow}
\hat n_{{\bf i},\downarrow} + \sum_{\bf i} \sum_{\sigma_1,\sigma_2}
V_{\sigma_1,\sigma_2} \hat n_{{\bf i},\sigma_1} \hat n_{{\bf i}+1,\sigma_2} , 
\nonumber\\
&+& \sum_{{\bf i},\sigma}(W_{1,\sigma} \hat n_{{\bf i},-\sigma} \hat c^{
\dagger}_{{\bf i},\sigma} \hat c_{{\bf i}+1,\sigma} + 
W_{2,\sigma} \hat n_{{\bf i}+1,-\sigma} \hat c^{\dagger}_{{\bf i}+1,\sigma} 
\hat c_{{\bf i},\sigma} + H.c.),
\label{eq1}
\end{eqnarray}
where $\hat c^{\dagger}_{{\bf i},\sigma}$ creates an electron with spin 
$\sigma$ at the site ${\bf i}$,
$\hat n_{{\bf i},\sigma}=\hat c^{\dagger}_{{\bf i},\sigma} \hat c_{
{\bf i},\sigma}$ represents the particle number operator, $t_{\sigma}$ is the
hopping matrix element, $U$ represents the Hubbard term, 
$V_{\sigma_1,\sigma_2}$ gives
the density-density nearest-neighbor interaction strengths, while
$W_{l,\sigma}$, $l=1,2$ describes the correlated hopping.

Inside the band, also a 
nearest neighbor spin-spin interaction is present given by
\begin{eqnarray}
\hat H_S = \sum_{\bf i} [ J_z \hat S^z_{\bf i} \hat S^z_{{\bf i}+1} +
\frac{J_{\perp}}{2} ( \hat S^{+}_{\bf i} \hat S^{-}_{{\bf i}+1} +
\hat S^{-}_{\bf i} \hat S^{+}_{{\bf i}+1} ) ],
\label{eq2}
\end{eqnarray}
where $\hat S^{\pm}=\hat S^x \pm i \hat S^y$, $\hat {\vec S}_{\bf i} =
\sum_{\alpha,\beta} \hat c^{\dagger}_{{\bf i},\alpha} {\vec \sigma}_{\alpha,
\beta} \hat c_{{\bf i},\beta}$, and ${\vec \sigma}_{\alpha,\beta}$ describes
the Pauli matrices. During this paper periodic
boundary conditions are used.


\section{The transformation of the Hamiltonian}

In order to deduce exact ground states for the Hamiltonian of the problem one
transforms $\hat H$ into an unique positive semidefinite form. In order to 
do this we introduce a bi-linear operator $\hat A_{\bf i}$ defined by
\begin{eqnarray}
\hat A_{\bf i} &=& z_1 \hat c^{\dagger}_{{\bf i}+1,\uparrow} \hat c^{\dagger}_{
{\bf i},\downarrow} + z_2 \hat c^{\dagger}_{{\bf i}+1,\downarrow} \hat c^{
\dagger}_{{\bf i},\uparrow} + z_3 \hat c^{\dagger}_{{\bf i}+1,\uparrow} 
\hat c^{\dagger}_{{\bf i}+1,\downarrow} + z_4 \hat c^{\dagger}_{{\bf i},
\uparrow} \hat c^{\dagger}_{{\bf i},\downarrow}
\nonumber\\
&+& w_1 \hat c_{{\bf i}+1,\uparrow} \hat c_{{\bf i},\downarrow} + 
w_2 \hat c_{{\bf i}+1,\downarrow} \hat c_{{\bf i},\uparrow} + 
w_3 \hat c_{{\bf i}+1,\uparrow} \hat c_{{\bf i}+1,\downarrow} + 
w_4 \hat c_{{\bf i},\uparrow} \hat c_{{\bf i},\downarrow},
\label{eq3}
\end{eqnarray}
where the prefactors $z_{\theta}, w_{\theta}$, $\theta=1,2,3,4$ 
are considered site independent numerical parameters.
One observes that $\sum_{\bf i} \hat A^{\dagger}_{\bf i} \hat A_{\bf i}$
reproduces the operators contained in Eqs.(\ref{eq1},\ref{eq2}), and
indeed one finds that the exact mapping
\begin{eqnarray}
\hat H = \sum_{\bf i} \hat A^{\dagger}_{\bf i} \hat A_{\bf i} -
a_{1} \hat N  - a_0
\label{eq4}
\end{eqnarray}
holds, if the numerical coefficients $z_{\theta}, w_{\theta}, a_{\sigma}, a_0$
are defined via the following matching conditions
\begin{eqnarray}
&&\frac{a_0}{N_{\Lambda}} = -a_{1}= \sum_{\theta=1}^4
|z_{\theta}|^2,
\nonumber\\
&&t_{\uparrow}=z^{*}_3 z_2 -z^{*}_1 z_4, \quad t_{\downarrow}=z^{*}_2 z_4 -
z^{*}_3 z_1 ,
\nonumber\\
&&V_{\uparrow,\uparrow} + \frac{J_z}{4} = V_{\downarrow,\downarrow} + 
\frac{J_z}{4} = 0,
\nonumber\\
&&V_{\uparrow,\downarrow} - \frac{J_z}{4}=|z_2|^2 + |w_2|^2, \quad
V_{\downarrow,\uparrow} - \frac{J_z}{4}=|z_1|^2 + |w_1|^2, 
\nonumber\\
&&U=|z_3|^2 + |z_4|^2 + |w_3|^2 + |w_4|^2 , \quad
\frac{J_{\perp}}{2}=w^{*}_1 w_2 + z^{*}_2 z_1,
\nonumber\\
&&w^{*}_3 w_1 + z^{*}_1 z_3 =W_{2,\downarrow}, \quad
w^{*}_3 w_2 + z^{*}_2 z_3 =-W_{2,\uparrow}, 
\nonumber\\ 
&&w^{*}_4 w_2 + z^{*}_2 z_4 =-W_{1,\downarrow}, \quad
w^{*}_4 w_1 + z^{*}_1 z_4 =W_{1,\uparrow},
\nonumber\\
&&w^{*}_3 w_4 + z^{*}_4 z_3 =F, \quad
w^{*}_1 z_2 + w^{*}_2 z_1 + w^{*}_3 z_4 + w^{*}_4 z_3 = 0 ,
\label{eq5}
\end{eqnarray}
where $N_{\Lambda}$ represents the number of lattice sites, and $\hat N$
is the total particle number operator. Since the total number of particles
is a constant of motion, via Eq.(\ref{eq4}) we matched the starting
Hamiltonian into an unique positive semidefinite form using the conditions
presented in Eq.(\ref{eq5}).


\section{The exact ground state at half filling}

Let us consider the bond operator
\begin{eqnarray}
\hat B^{\dagger}_{\bf i} = \alpha_{\uparrow} \hat c^{\dagger}_{{\bf i},
\uparrow} + \alpha_{\downarrow} \hat c^{\dagger}_{{\bf i},\downarrow} + 
\beta_{\uparrow} \hat c^{\dagger}_{{\bf i}+1,\uparrow} + 
\beta_{\downarrow} \hat c^{\dagger}_{{\bf i}+1,\downarrow},
\label{eq6}
\end{eqnarray}
where $\alpha_{\sigma}$ and $\beta_{\sigma}$ are numerical prefactors.
One observes that $\{ \hat B^{\dagger}_{\bf i}, \hat B^{\dagger}_{\bf j} \}=0$
holds for all ${\bf i},{\bf j}$ and all $\alpha_{\sigma}, \beta_{\sigma}$.

Let us analyze the wave function
\begin{eqnarray}
|\Psi_g\rangle = \prod_{{\bf i}=1}^{N_{\Lambda}} \hat B^{\dagger}_{\bf i} 
|0\rangle,
\label{eq7}
\end{eqnarray}
where $|0\rangle$ is the bare vacuum with no fermions present. Since in 
Eq.(\ref{eq7}) one introduces $N_{\Lambda}$ electrons into the system, 
$|\Psi_g\rangle$ is defined at half filling.

If for all ${\bf i}$ one has 
\begin{eqnarray}
\hat A_{\bf i} |\Psi_g\rangle = 0,
\label{eq8}
\end{eqnarray}
then clearly, $|\Psi_g\rangle$ represents the ground state wave function of
$\hat H$. Since the fermionic operators from $\hat A_{\bf i}$ in Eq.(\ref{eq3})
are present in $\prod_{\bf j} \hat B^{\dagger}_{\bf j}$
only in $\hat B^{\dagger}_{{\bf i}}, \hat B^{\dagger}_{{\bf i}\pm 1}$,
it results that in order to satisfy Eq.(\ref{eq8}) one must has for all 
${\bf i}$ sites
\begin{eqnarray}
\hat A_{\bf i} \hat B^{\dagger}_{\bf i} [\hat B^{\dagger}_{{\bf i}-1} 
\hat B^{\dagger}_{{\bf i}+1} ] |0\rangle = 0 ,
\label{eq9}
\end{eqnarray}
from where one finds
\begin{eqnarray}
&&z_1\beta_{\downarrow} - z_3 \alpha_{\downarrow} =0, \quad
z_2 \beta_{\uparrow} + z_3 \alpha_{\uparrow} =0, \quad
z_1\alpha_{\uparrow} - z_4 \beta_{\uparrow} =0, \quad
z_2\alpha_{\downarrow} + z_4 \beta_{\downarrow} =0, 
\nonumber\\
&&w_1 (\alpha_{\uparrow}^2\beta_{\downarrow} - \alpha_{\uparrow} \alpha_{
\downarrow} \beta_{\uparrow}) + w_3(\beta^2_{\uparrow} \alpha_{\downarrow} -
\beta_{\uparrow}\beta_{\downarrow}\alpha_{\uparrow}) =0,
\nonumber\\
&&w_2 (\alpha_{\downarrow}^2\beta_{\uparrow} - \alpha_{\uparrow} \alpha_{
\downarrow} \beta_{\downarrow}) - w_3(\beta^2_{\downarrow} \alpha_{\uparrow} -
\beta_{\uparrow}\beta_{\downarrow}\alpha_{\downarrow}) =0,
\nonumber\\
&&w_4 (\alpha_{\downarrow}^2\beta_{\uparrow} - \alpha_{\uparrow} \alpha_{
\downarrow} \beta_{\downarrow}) + w_1(\beta^2_{\downarrow} \alpha_{\uparrow} -
\beta_{\uparrow}\beta_{\downarrow}\alpha_{\downarrow}) =0,
\nonumber\\
&&-w_4 (\alpha_{\uparrow}^2\beta_{\downarrow} - \alpha_{\uparrow} \alpha_{
\downarrow} \beta_{\uparrow}) + w_2(\beta^2_{\uparrow} \alpha_{\downarrow} -
\beta_{\uparrow}\beta_{\downarrow}\alpha_{\uparrow}) =0,
\nonumber\\
&&-w_4 (\beta_{\downarrow}^2\alpha_{\uparrow} - \beta_{\uparrow} \beta_{
\downarrow} \alpha_{\downarrow}) - w_1\beta^2_{\downarrow} \beta_{\uparrow} -
w_2 \beta^2_{\downarrow}\beta_{\uparrow} =0,
\nonumber\\
&&w_4 (\beta_{\uparrow}^2\alpha_{\downarrow} - \beta_{\uparrow} \beta_{
\downarrow} \alpha_{\uparrow}) - w_1\beta^2_{\uparrow} \beta_{\downarrow} -
w_2 \beta^2_{\uparrow}\beta_{\downarrow} =0,
\nonumber\\
&&w_3 (\alpha_{\downarrow}^2\beta_{\uparrow} - \alpha_{\uparrow} \alpha_{
\downarrow} \beta_{\downarrow}) - w_1\alpha^2_{\downarrow} \alpha_{\uparrow} -
w_2 \alpha^2_{\downarrow}\alpha_{\uparrow} =0,
\nonumber\\
&&-w_3 (\alpha_{\uparrow}^2\beta_{\downarrow} - \alpha_{\uparrow} \alpha_{
\downarrow} \beta_{\uparrow}) - w_1\alpha^2_{\uparrow} \alpha_{\downarrow} -
w_2 \alpha^2_{\uparrow}\alpha_{\downarrow} =0.
\label{eq10}
\end{eqnarray}
In Eq.(\ref{eq10}) the first row comes from the coefficients of the
terms containing three creation operators created by the product 
$\hat A_{\bf i} \hat B^{\dagger}_{\bf i}$, while the remaining equations are
provided by the rest of the terms from Eq.(\ref{eq9}). 

Based on Eqs.(\ref{eq4},\ref{eq8}), the ground state energy is given by
$E_g= -a_0-a_1 N$.

\section{The deduced solutions}

In the following one presents the obtained solutions for the system of 
equations Eqs.(\ref{eq5},\ref{eq10}) denoted by Solutions I. and II. 
which describe two different regions of the phase diagram.

\subsection{Solution I.}

Taking into consideration $W_{1,\sigma}=W_{2,\sigma}=W$, the parameters of
the $\hat A_{\bf i}, \hat B_{\bf i}$ operators become
\begin{eqnarray}
&&z_2=-z_1, \quad z_3=q z_1 e^{i\alpha}, \quad z_4=\frac{z_1}{q}e^{-i\alpha},
\quad w_1=\frac{z_1}{2}(\frac{|q|^2-1}{q}),
\nonumber\\
&&w_2=-w_1, \quad w_3=-\frac{z_1 q^{*}}{q} e^{-i\alpha}, \quad
w_4=z_1 e^{i\alpha},
\nonumber\\
&&\alpha_{\uparrow}=\lambda \alpha_{\downarrow}, \quad \beta_{\uparrow}=
\lambda q \alpha_{
\downarrow} e^{i\alpha}, \quad \beta_{\downarrow}=q \alpha_{\downarrow}
e^{i\alpha},
\label{eq11}
\end{eqnarray}
where $z_1, \alpha, \alpha_{\downarrow}, \lambda, q$ remain 
arbitrary coefficients
at the level of (\ref{eq11}). The Hamiltonian parameters turn to have the 
property $t=t_{\sigma}$, $V_1=V_{\sigma,\sigma}$, $V_2=V_{\sigma,-\sigma}$,
$a_0=U N_{\Lambda}$, $a_1=-U$, and one has
\begin{eqnarray}
W=-\frac{t}{2}, \quad V_1=-\frac{J_z}{4}, \quad V_2=\frac{U+J_z}{4}, \quad
J_{\perp}=-\frac{U}{2},
\label{eq12}
\end{eqnarray}
where $U >0$, $t$ and $J_z$ can be arbitrary chosen, and the parameters 
$|z_1|,|q|,
\alpha$ becomes determined by $|z_1|=|t|/\sqrt{U}$, $(1+|q|^2)/|q|=|t|/|z_1|^2
$, and $e^{-i\alpha}= (q/|q|) sign(t)$. The coefficients $\lambda$ and 
$\alpha_{\downarrow}$ remain arbitrary, and the operator constructing 
the ground state wave function from (\ref{eq7}) becomes
\begin{eqnarray}
\hat B^{\dagger}_{\bf i}= \alpha_{\downarrow}[(\hat c^{\dagger}_{{\bf i},
\downarrow} + \lambda \hat c^{\dagger}_{{\bf i},\uparrow})+ q e^{i\alpha}
(\hat c^{\dagger}_{{\bf i}+1,\downarrow} + \lambda \hat c^{\dagger}_{{\bf i}+1,
\uparrow})] .
\label{eq13}
\end{eqnarray}

\subsection{Solution II.} 

If one considers $W_{1,\sigma}=W_1$, $W_{2,\sigma}=W_2$ and takes
$\nu=z_3/w_4=-w^*_3/z^*_4$ non-zero and real, the parameters of $\hat A_{
\bf i}, \: \hat B_{\bf i}$ operators become

\begin{eqnarray}
&&z_2=-z_1, \quad z_3= P e^{i\alpha} z_1, \quad 
z_4= \frac{e^{-i\alpha}}{P} z_1,
\quad w_1=-z_1, 
\nonumber\\
&&w_2=z_1, \quad
w_3=-\nu \frac{e^{-i\alpha}}{P}z_1, \quad w_4=\frac{1}{\nu} P e^{i\alpha} 
z_1, 
\nonumber\\
&&\alpha_{\uparrow}=\lambda \alpha_{\downarrow}, \quad
\beta_{\uparrow} = \lambda P e^{i\alpha} \alpha_{\downarrow}, \quad
\beta_{\downarrow}= P e^{i\alpha} \alpha_{\downarrow},
\label{eq14}
\end{eqnarray}
where $P=\sqrt{|\nu|/(\sqrt{2} + sign(\nu))}$, $\alpha$ is given by
$sign(t)=-e^{-i\alpha}$,
and $z_1,\alpha_{\downarrow},\lambda,\nu$ are arbitrary non-zero coefficients.
Furthermore one has $a_1=a_0/N_{\Lambda}=\sqrt{W_1W_2/2}
(2+P^2+P^{-2})$, and the Hamiltonian parameters become
\begin{eqnarray}
&&W_1=-sign(t) \frac{\sqrt{2}|z_1|^2}{|\nu|} \sqrt{P}, \quad
W_2=-sign(t)\sqrt{2}|z_1|^2 |\nu| \sqrt{P^{-1}},
\nonumber\\
&&J_{\perp}=-\sqrt{2W_1W_2}, \quad V_1=-\frac{J_z}{4}, \quad V_2=\frac{J_z}{4}
-J_{\perp}, 
\nonumber\\
&&t=-\frac{1}{\sqrt{2}} (\frac{W_2}{|\nu|} + |\nu|W_1), \quad
U=\frac{1+|\nu|^2}{|\nu| \sqrt{2W_1W_2}} (|\nu|W_1^2 + \frac{W_2^2}{|\nu|}),
\label{eq15}
\end{eqnarray}
where $\nu,J_z$ remain arbitrary, $sign(W_1)=sign(W_2)$ must hold, and one has
$|z_1|^2=\sqrt{W_1W_2/2}$. The operator present in the expression of the
ground state wave function (\ref{eq7}) is given by
\begin{eqnarray}
\hat B_{\bf i}^{\dagger}=\alpha_{\downarrow} [ (\lambda \hat c^{\dagger}_{
{\bf i},\uparrow} + \hat c^{\dagger}_{{\bf i},\downarrow}) - P sign(t)
(\lambda \hat c^{\dagger}_{{\bf i}+1,\uparrow} + \hat c^{\dagger}_{{\bf i}+1,
\downarrow}) ],
\label{eq16}
\end{eqnarray}
so its form is similar to that given in Eq.(\ref{eq13}).

\subsection{Physical properties of the deduced solutions} 

For the ground state (\ref{eq7}), in both obtained cases 
(\ref{eq13},\ref{eq16})
describing different regions of the phase diagram, the 
$\hat B^{\dagger}_{\bf i}$ operator is constructed from 
$\hat I^{\dagger}_{\bf i} = (\lambda \hat c^{\dagger}_{{\bf j},\uparrow} + 
\hat c^{\dagger}_{{\bf j},\downarrow})$ type of blocks, for which
$\hat I^{\dagger}_{\bf j} \hat I^{\dagger}_{\bf j} =0$ holds for arbitrary
$\lambda$. Consequently, apart from a normalization constant,
$\prod_{{\bf i}=1}^{N_{\Lambda}} \hat B^{\dagger}_{\bf i} |0\rangle$ from
(\ref{eq7}) becomes $\prod_{{\bf i}=1}^{N_{\Lambda}} \hat I^{\dagger}_{\bf i}
|0\rangle$, hence the normalized ground state in both cases of Solutions I. 
and II. can be written as
\begin{eqnarray}
|\Psi_g\rangle = (1+|\lambda|^2)^{-N_{\Lambda}/2} \prod_{{\bf i}=1}^{N_{
\Lambda}}(\lambda \hat c^{\dagger}_{{\bf j},\uparrow} + 
\hat c^{\dagger}_{{\bf j},\downarrow})|0\rangle,
\label{eq17}
\end{eqnarray}
where $\lambda$ is arbitrary, and $N=N_{\Lambda}$ in (\ref{eq17}) fixes the 
filling at half. Denoting by $\langle ... \rangle = \langle 
\Psi_g | ... |\Psi_g\rangle$ ground state expectation values, and using
as usual
$\hat S^z_{\bf i}=1/2(\hat n_{{\bf i},\uparrow}-\hat n_{{\bf i},\downarrow})$,
$\hat S^{+}_{\bf i}= (\hat S^{-})^{\dagger}_{\bf i} = \hat c^{\dagger}_{
{\bf i},\uparrow} \hat c_{{\bf i},\downarrow}$, $\hat S^{\pm}_{\bf i} =
\hat S^{x}_{\bf i} \pm i \hat S^{y}_{\bf i}$, and $\hat {\vec S} = \sum_{\bf i}
\hat {\vec S_{\bf i}}$ for the spin operators, 
one finds $\langle \hat S^2 \rangle = (N_{\Lambda}/2)(
N_{\Lambda}/2 +1)$. This result is motivated by the fact that since 
$\lambda$ is the same on each ${\bf i}$, the local spin orientations are the 
same on each site. Hence the ground state represents a fully saturated 
ferromagnetic state. Furthermore, since only local Hamiltonian terms contribute
in the ground state energy, and independent on ${\bf i}$,
$\langle \hat n_{{\bf i},\uparrow} \rangle =1- \langle \hat n_{{\bf i},
\downarrow} \rangle = |\lambda|^2/(1+|\lambda|^2)$, 
$\langle \hat n_{{\bf i},\uparrow} \hat n_{{\bf i},\downarrow} \rangle=0$, 
the ground state is 
localized \cite{int4}. Consequently, the obtained ground state at half 
filling of the studied one (dispersive) band model is a localized and fully 
polarized ferromagnet.
 
Note the important aspect that $|\Psi_g\rangle$ is the ground state only of 
the full Hamiltonian. Hence not a specific band structure produces here the 
ferromagnetic behaviour (as for example in the case of the flat band 
ferromagnetism \cite{f1}, or Lieb's degenerate middle band magnetism in a 
bipartite lattice \cite{f2}), nor the interactions select the ground state
from the minimum energy eigenstates of the kinetic part \cite{f2a,f3}.
In the described case the ferromagnetism at half
filling can appear in exact terms in an arbitrary dispersive band being 
only the ground state of the full Hamiltonian, e.g. is provided by an 
incontestable correlation effect. 

As described previously in details \cite{int4,int6}, for a non-integrable 
model as that studied here, exact ground states can be obtained only on
different cuts of the phase diagram corresponding to different decompositions
in positive semidefinite form of the starting Hamiltonian. Since for the 
presented two solutions in Sections V.A, V.B, the expression of the operator 
$\hat A_{\bf i}$ is different, these solutions correspond to different 
decompositions, hence provide the deduced ground state in different regions
of the phase diagram. The fact that the same solution appears in different
regions underlines the stability of the ferromagnetic phase, since
finite $\hat H$ parameter modifications lead back to the same ordered phase.
However, the ferromagnetic localized phase has a finite extension. This is seen
from the fact that since $|\Psi_g\rangle$ is not an eigenstate
of the non-interacting Hamiltonian, turning off the interactions one
recovers the conducting behaviour of a non-interacting dispersive band, e.g. a 
paramagnetic metal to ferromagnetic insulator transition must be present in
the phase diagram of the system.  

We further note that resembling ferromagnetic states have been found in
similar systems by other methods as well \cite{uj1,uj2,uj3}.

\section{Summary and Conclusions}

Single band extended Hubbard models containing spin-spin interactions 
are studied by a
positive semidefinite operator technique, leading to ferromagnetic and
localized exact ground states in different regions of the phase diagram via 
quadratic operators used in the positive semidefinite decomposition
leading to an unique operator term in $\hat H$. Hence the deduced
ground states are  minimum energy eigenstate only of the full 
Hamiltonian, e.g. not are connected to a specific band structure, nor are
separately eigenstates of the kinetic and interacting parts.
The ferromagnetism emerges at half filling in an arbitrary
dispersive band, being provided by clear correlation effects.

\acknowledgements

We kindly acknowledge financial support of the grant OTKA-T48782 of 
Hungarian Scientific Research Fund, and in case of Zs.G. also from Alexander 
von Humboldt Foundation.



\begin{thebibliography}{99}
\bibitem{f2a}
We include here also bond-charge terms in the kinetic part of $\hat H$.
\bibitem{f3}
This is the case when several positive semidefinite operators are used in 
the expression of the transformed Hamiltonian.
\end{thebibliography}
\end{document}